\documentclass[%
aps,
reprint,
groupedaddress,
 amsmath,amssymb,
 prappl,
]{revtex4-2}

\usepackage{graphicx}
\usepackage{dcolumn}
\usepackage{multirow}
\usepackage{bm}
\usepackage{siunitx}
\usepackage{braket}
\usepackage{amsmath}
\usepackage{hyperref}
\usepackage{soul}
\usepackage[english,italian]{babel}
\renewcommand{\selectlanguage}[1]{}
\DeclareSIUnit\dBm{dBm}
\DeclareSIUnit{\rad}{rad}

\begin{document}


\title{\textbf{Observation and mitigation of microwave echoes from dielectric defects in Josephson traveling wave amplifiers} 
}

\author{Matteo Boselli}%
\affiliation{\textsc{CNRS}$,$ \textsc{ENS de Lyon}$,$ \textsc{LPENSL}$,$ \textsc{UMR5672}$,$ \textsc{69342}~\text{Lyon~cedex~07}$,$ \text{France}}

\author{Joel Grebel}%
\thanks{Present address: Google, Santa Barbara, CA, USA}
\affiliation{\textsc{CNRS}$,$ \textsc{ENS de Lyon}$,$ \textsc{LPENSL}$,$ \textsc{UMR5672}$,$ \textsc{69342}~\text{Lyon~cedex~07}$,$ \text{France}}

\author{Ambroise Peugeot}%
\affiliation{\textsc{CNRS}$,$ \textsc{ENS de Lyon}$,$ \textsc{LPENSL}$,$ \textsc{UMR5672}$,$ \textsc{69342}~\text{Lyon~cedex~07}$,$ \text{France}}

\author{Rémy Dassonneville}%
\thanks{Present address: \textsc{CNRS}$,$ \textsc{Aix-Marseille Univ.}$,$ \textsc{University of Toulon}$,$ \textsc{IM2NP}$,$ \textsc{UMR7334}$,$ \textsc{13013}~\text{~Marseille}$,$ \text{France}}
\affiliation{\textsc{CNRS}$,$ \textsc{ENS de Lyon}$,$ \textsc{LPENSL}$,$ \textsc{UMR5672}$,$ \textsc{69342}~\text{Lyon~cedex~07}$,$ \text{France}}

\author{Benjamin Huard}%
\affiliation{\textsc{CNRS}$,$ \textsc{ENS de Lyon}$,$ \textsc{LPENSL}$,$ \textsc{UMR5672}$,$ \textsc{69342}~\text{Lyon~cedex~07}$,$ \text{France}}

\author{Audrey Bienfait}%
\email{audrey.bienfait@ens-lyon.fr}
\affiliation{\textsc{CNRS}$,$ \textsc{ENS de Lyon}$,$ \textsc{LPENSL}$,$ \textsc{UMR5672}$,$ \textsc{69342}~\text{Lyon~cedex~07}$,$ \text{France}}

\date{\today}

\begin{abstract}

Amplifying microwave signals with a noise close to the minimum imposed by quantum mechanics is now routinely performed with superconducting quantum devices. In particular, Josephson-based Traveling Wave Parametric Amplifiers (JTWPA) have shown record bandwidth with added noise close to the quantum limit~\cite{macklin2015,planat2020}. In this work, we report the appearance of echo signals emitted by JTWPAs driven by trains of high-power pulses near or exceeding their dynamical range.
These echoes have micro-second coherence and we attribute their origin to microscopic defects in the amplifier dielectric layer. By analyzing the power and the coherence of the echo signal as a function of temperature, we estimate the dielectric loss brought by these defects, and their impact on the JTWPA quantum efficiency. We introduce a mitigation technique (BLAST) to prevent the appearance of these echoes, which can alter measurements in experiments. It consists in an additional off-resonant high-power tone sent concurrently with each pulse. We demonstrate that it suppresses the spurious defect signals and we recover the typical gain and noise figure within 95 $\%$ of their low-power values in $\qty{300}{\nano\second}$. These results can help to extend the use of JTWPAs in experiments where fast high-power sequences are necessary to generate weak microwave responses from the system under study, and also provide a path towards characterizing in-situ the dielectric losses of these devices.\\
\end{abstract}


\maketitle

Parametric amplification is now a widespread technique for efficiently measuring quantum microwave states~\cite{clerk2010} and generating squeezed states~\cite{bergeal2010,grimsmo2017,esposito2022,qiu2023}.
It is thus key for high-fidelity readout of qubits~\cite{vijay2011,stehlik2015} and for microwave quantum sensing~\cite{teufel2011,smith2013,bockstiegel2014,bienfait2015}. Many amplifier designs have been explored, among which traveling-wave amplifiers based on Josephson junctions (JTWPAs)~\cite{macklin2015,planat2020} have demonstrated large gain over bandwidths exceeding \num{2}~\unit{\giga\hertz}, with added noise close to the quantum limit. These performances are reached when the signal strength is kept below the \num{1}~\unit{\decibel} compression point of these devices, about $\qty{-100}{\dBm}$ for current devices~\cite{macklin2015,planat2020}. 
This is a limitation when one would like to measure a signal along with a drive pulse exceeding this \num{1}~\unit{\decibel} compression point. Here, we show that the limitation also arises when the drive pulses are applied less than a few microseconds before the signal to be detected and also enter the JTWPA. This is typically the case when strong drive pulses are required to trigger a response from the system under study such as performing fast gates and readout for qubits~\cite{krantz2019}, and pulse-probe experiments~\cite{albanese2020,bartram2023,casariego2023,youssefi2023,rej2024}. Here, we probe the dynamics of JTWPAs provided by the Lincoln labs~\cite{macklin2015}. 

Specifically, we observe the emergence of a delayed spurious signal in response to high-power pulses that we attribute to the collective coherent excitation and emission of dielectric defects inside the JTWPA capacitors. This feature is dependent on temperature and vanishes for $T>\qty{110}{\milli\kelvin}$, similarly to what is expected from dielectric echoes observed in spin glass physics~\cite{golding1976,black1977,bernard1978}. By studying the dynamics of this effect, we quantify an associated microwave absorption limiting the efficiency of the JTWPA~\cite{macklin2015,houde2019,yuan2022}. To counteract the in-situ emission of these spurious signals, we propose and demonstrate two mitigation strategies. One strategy relies on interferometric cancellation of the drive pulses at the JTWPA input. The other consists in adding, on top of the drive pulses, an off-resonant pulse whose power is well above the 1~dB compression point of the JTWPA and of all other drive pulses. This pulse, that we call BLAST (BLinding for Amplification Suppression Technique), turns the JTWPA into a fully reflective device, preventing any signal from exciting the microscopic defects, thus avoiding the emission of the spurious echo. We observe that a transient period of $\qty{300}{\nano\second}$ is necessary to fully recover the low-power behavior of the JTWPA after these BLAST pulses, a timing similar to the transient time needed to stabilize the JTWPA gain when powered on.

\section{Experimental setup}

The JTWPAs we use are made up of a chain of Josephson junctions connected in series, and grounded through parallel-plate capacitors. The dielectric of these capacitors is a SiO$_2$ and NbO$_x$ bilayer~\cite{macklin2015}. Every three Josephson junctions, an LC resonator is added to ensure quasi-phase matching between a signal at frequency $\omega_\mathrm{d}$ and a pump tone at frequency $\omega_\mathrm{p}$, ensuring that a four-wave amplification process takes place constructively all along the chain, yielding a gain of about 20~dB over a bandwidth larger than $\qty{2}{\giga\hertz}$ when the amplification pump tone is present. These resonators also create a dispersive gap which prevents transmission through the JTWPA over $\sim\qty{100}{\mega\hertz}$ band at the center of the frequency gain profile. 

\begin{figure}[h]
    \centering
    \includegraphics{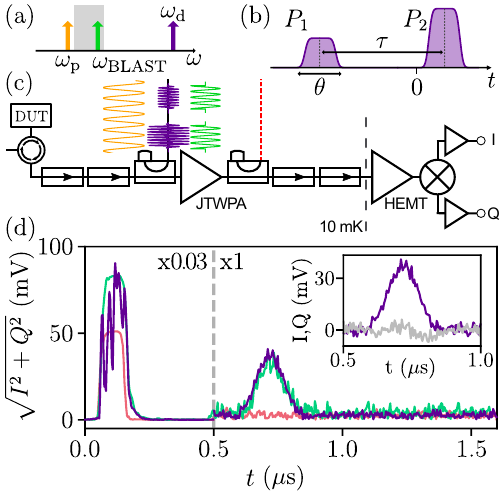}
    \caption{(a) Frequencies of the experimental tones compared to the dispersive feature of the JTWPA (gray): JTWPA pump tone at $\omega_{\mathrm{p}}$ (orange), BLAST tone at $\omega_{\mathrm{BLAST}}$ (green), driving tone at $\omega_{\mathrm{d}}$ (purple). (b) Two pulse sequence. (c) The pulses are directed either at the input of the JTWPA (purple) or at the input of a HEMT amplifier (red) using directional couplers. The transmitted signals are amplified by room-temperature amplifiers and demodulated. (d) Detected signal amplitude on JTWPA D1 when routing the two pulses through the JTWPA with the JTWPA pump tone (light green) or without it (purple) and through the HEMT alone (light red). We operate at $\omega_{\mathrm{d}}/(2\pi)=\qty{7}{\giga\hertz}$ and signal powers can be found in the text. All traces show the transmitted $P_2$ pulse at $t\sim\qty{0.1}{\micro\second}$ but a spurious signal is present at a time $\tau=\qty{0.6}{\micro\second}$ later only when going through the JTWPA. We have applied a scaling factor $\times 0.03$ during the first period of the trace. In the inset, I (purple) and Q (gray) quadratures of the spurious signal detected with no JTWPA pump.}
     \label{fig.1}
\end{figure}

The JTWPA is inserted in a typical setup for quantum microwave experiments (Fig \ref{fig.1}c and Appendix \ref{supp:fridge_setup}). It is anchored at the base plate of a dilution refrigerator, whose temperature $T$ is varied from $\qty{8}{\milli\kelvin}$ to $\qty{110}{\milli\kelvin}$. Any signal emitted from a device under test (DUT) enters the JTWPA after combination with the amplification pump tone using a directional coupler. After the JTWPA, the signal is routed to a high electron mobility transistor (HEMT) cryogenic amplifier through two isolators, followed by room-temperature amplification and heterodyne demodulation, allowing to access the $I$ and $Q$ quadratures of the signal. 

\section{Observation of two-pulse echoes}

We use a drive sequence at frequency $\omega_{\mathrm{d}}$ comprising two pulses $P_1$ and $P_2$ of length $\theta$ separated by an interval $\tau$ (Fig. \ref{fig.1}b), akin to a Hahn echo experiment in magnetic resonance. The drive can be at any frequency within the $\num{4}$-$\qty{8}{\giga\hertz}$ bandwidth of the isolators, whereas the pump and BLAST tones lie close to the JTWPA dispersive feature (see Fig.~\ref{fig.1}a). This drive sequence is applied through the JTWPA pump line, bypassing entirely the DUT. Unless specified otherwise, each pulse is a flat-top pulse of length $\theta=\qty{100}{\nano\second}$ with a rise time of $\qty{20}{\nano\second}$ and we do not apply any pump tone to activate the JTWPA amplification. In Fig.~\ref{fig.1}d, we observe that when this two-pulse sequence is sent to the JTWPA, it triggers the emission of a spurious signal of Gaussian shape. The powers of the pulses going through the JTWPA when no pump tone is applied (purple) are set to $P_1=\qty{-81}{\dBm}$ and $P_2=\qty{-75}{\dBm}$ referred to the JTWPA input. To check whether this signal is solely due to the JTWPA, we reroute this two-pulse drive directly through the HEMT. Even using higher powers ($P_{1(2)}^{\mathrm{max}} = \qty{-50}{\dBm}$ at the HEMT input), we do not observe this spurious signal. Conversely, when going through the JTWPA with the amplification tone turned on (with gain $G=\qty{19.7}{\dB}$), we observe an echo of the same amplitude (green) at considerably lower powers ($P_1=\qty{-98}{\dBm}$ and $P_2=\qty{-92}{\dBm}$ at the JTWPA input). These power levels are similar to what is used when performing qubit readout and gates, so that the spurious signal may interfere with experiments as seen in~\cite{dassonneville2023}. Since the spurious signal is dominantly emitted on a single quadrature (see inset of Fig.~\ref{fig.1}d), in the following we post-process all signals to only keep this quadrature of interest $I$, and compute a mean average signal $\bar{I}$ (see Appendix~\ref{supp:fitEchoRoutine}). 

\begin{figure}[h]
	\includegraphics{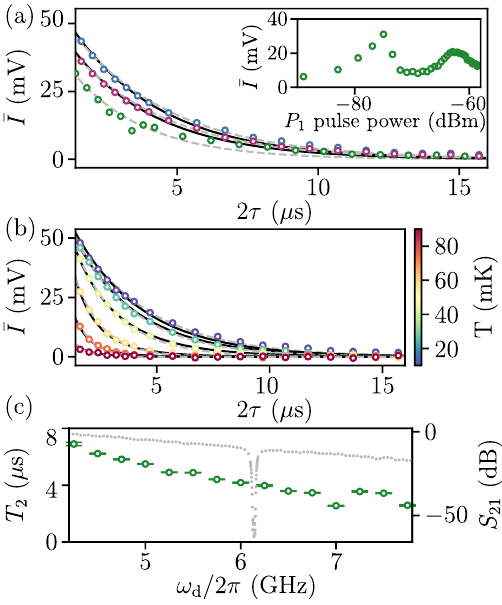}
    \caption{(a) Mean echo amplitude $\bar{I}$ versus delay $2\tau$ between $P_1$ and the echo. The measurements are performed at temperature $T=\qty{8}{\milli\kelvin}$ and $\omega_{\mathrm{d}}/(2\pi)=\qty{7.0}{\giga\hertz}$ on amplifier D1 (green), D2 (blue) and D3 (pink). Lines are fit either using a simple exponential model (gray dashes) or using the spectral diffusion model discussed in the main text (black). In inset, average echo amplitude  $\bar{I}$ versus $P_1$ pulse peak power, with $P_2=\qty{-77}{\dBm}$ for JTWPA D1. (b) Mean echo amplitude $\bar{I}$ versus delay $2\tau$  measured on device D3 at $T=\qty{10}{\milli\kelvin}$, $\qty{18}{\milli\kelvin}$, $\qty{26}{\milli\kelvin}$, $\qty{40}{\milli\kelvin}$, $\qty{60}{\milli\kelvin}$ and $\qty{90}{\milli\kelvin}$. Gray dotted lines, simple exponential fit. Black lines, spectral diffusion model. (c) Hahn echo decay time $T_2$ as a function of $\omega_{\mathrm{d}}$. Gray dots: measured transmission amplitude profile through amplifier D1 including the contribution of the lines}
    \label{fig.2}
\end{figure}

To identify the origin of these spurious signals, we explore different pulse powers, delays, and JTWPA devices. We first note that we observe such signals on three different amplifiers $\mathrm{D}1$, $\mathrm{D}2$, $\mathrm{D}3$ that respectively have dispersive features centered on $\qty{6.14}{\giga\hertz}$, $\qty{6.12}{\giga\hertz}$ and $\qty{7.94}{\giga\hertz}$. 
When sweeping the delay $\tau$ between pulses, we see a decay of the integrated signal (see Fig.~\ref{fig.2}a).  We first fit this decay with a simple exponential $A_0 e^{-(2\tau/T_2)}$, extracting coherence times $T_{2}^{\mathrm{D}1}=\qty{2.55\pm0.02}{\micro\second}$, $T_{2}^{\mathrm{D}2}=\qty{3.50\pm0.02}{\micro\second}$ and $T_{2}^{\mathrm{D}3}=\qty{3.42\pm0.02}{\micro\second}$ depending on the device, for data acquired at $T=\qty{8}{\milli\kelvin}$. When increasing the temperature we observe that the coherence time is reduced (see Fig.~\ref{fig.2}b): for instance, $T_{2}^{\mathrm{D}3}=\qty{0.61\pm0.03}{\micro\second}$ at $T=\qty{90}{\milli\kelvin}$. The amplitude of the signal also depends on the power of the pulses. While keeping $P_2$ fixed in inset of Fig.~\ref{fig.2}a, we observe that $\bar{I}$ oscillates as a function of $P_1$ as in a Rabi experiment for one or several two-level systems (TLS). We only observe up to one or two oscillations, indicating either a very short Rabi coherence time or a large spread in coupling strengths between the field and the TLSs responding to the sequence~\cite{sigillito2014}. In all devices, we observe the presence of this signal in the \num{4} to $\qty{8}{\giga\hertz}$ frequency range accessible in our setup, at frequencies both above and below the JTWPA dispersive features, with coherence decreasing for higher $\omega_\mathrm{d}$ (see Fig.~\ref{fig.2}c).

\section{Modeling: ensemble of microscopic dielectric defects}
From these experimental observations, we attribute the physical origin of the echo signal to microscopic defects in the JTWPA dielectric layer. Similar signals have already been observed in dielectrics (also called spin glass) at low-temperature since the 1960s, using acoustic~\cite{golding1976,black1977} or microwave drives~\cite{vonschickfus1977,bernard1978,baier1988}. These signals were first explored to explain the anomalous heat coefficient of amorphous dielectrics at low temperature, and attributed to microscopic defects. To model these defects and their resulting echoes, the standard tunneling model was put forward~\cite{black1977}. Later on, this model was extended to explain the absorption of microwave radiation by microscopic defects~\cite{burin2013,burin2015,muller2019}, known to limit the performances of superconducting resonators and qubits~\cite{kjaergaard2020}. In the context of parametric amplification, beyond limiting the quantum efficiency, TLS have been proposed to explain the power behavior of intermodulation products~\cite{kaufman2023}.

A single defect can be modeled as a two-level-system with Hamiltonian $H/\hbar=\frac{\Delta}{2}\sigma_z + \frac{\Delta_0}{2} \sigma_x$ where $\Delta$ is the energy splitting between the two levels, $\Delta_0$ the tunneling barrier between the two states and $\sigma_{x(z)}$ are the  $x(z)$ Pauli matrices. From this model, an ensemble of TLSs responds to the same Bloch equations as an ensemble of spins and is thus expected to produce echoes when probed with Hahn echo sequences as in Fig.~\ref{fig.1}b~\cite{black1977}. In a material, a wide distribution of TLS resonant frequencies ($\omega_0 = \sqrt{\Delta_0^2 + \Delta^2})$ is expected ranging from zero frequency to some finite cut-off value which depends on the specific nature of the TLS and the properties of the host dielectric~\cite{phillips1972,black1977}.
It would thus fit well with our observations over the entire \num{4}-$\qty{8}{\giga\hertz}$ band.

When probing the ensemble of TLSs of frequency $\omega_{\mathrm{A}}$ that are resonant with the drive $\omega_{\mathrm{d}}=\omega_{\mathrm{A}}$ (TLSs A), we expect the TLSs to be fully polarized at the temperature $T\ll\hbar\omega_{\mathrm{d}}/k_\mathrm{B}$ we operate at. However, lower frequency TLSs (TLSs B) are expected to be thermally excited. TLS-TLS dipolar interaction couples TLSs B to TLSs A. When TLSs B flip randomly, they induce uncontrollable frequency shifts on the TLSs A excited by the initial pulse $P_1$ resulting in broadening the spectral distribution of excited TLSs A. This decoherence mechanism, known as spectral diffusion~\cite{mims1961,black1977}, is strongly temperature dependent. Among spectral diffusion models~\cite{mims1961,klauder1962} we make use of an uncorrelated jump model~\cite{hu1974} to capture the temperature dependence of the decoherence (see Appendix \ref{supp:Spectral_Model}). In this model, TLSs A are assumed identical and similarly coupled to the bath of identical thermally excited TLSs B. The dipolar diffusion rate is given by:
\begin{equation}\label{eqn.gammaSD}
\Gamma_{\mathrm{sd}}(T)=\frac{2\pi}{9\sqrt{3}\hbar\epsilon}d_{\mathrm{A}}d_{\mathrm{B}} c_{\mathrm{B}} \mathrm{sech}^2(\hbar \omega_{\mathrm{B}}/(2k_\mathrm{B}T))
\end{equation}
where $\epsilon$ is the permittivity of the dielectric material, $d_{\mathrm{A}}$,  $d_{\mathrm{B}}$ are the dipole moments of TLSs A and B, $\omega_{\mathrm{B}}$ and $c_{\mathrm{B}}$ are the TLSs B frequency and concentration, expressed as number of TLSs per unit of volume.
The hyperbolic secant term expresses that only B TLSs that have flipped participate to spectral diffusion: at low temperature ($
T\ll\hbar \omega_{\mathrm{B}}/k_{\mathrm{B}}$) the process is frozen, while it reaches its maximum rate $\Gamma_{\mathrm{sd}}^0=\Gamma_{\mathrm{sd}}(T\rightarrow\infty)$ at large temperature. The Hahn echo amplitude measured at temperature $T$ is given by:
\begin{equation}
A(2\tau,T)=A_0(T) e^{-2\Gamma_2\tau} e^{-\Gamma_{\mathrm{sd}}(T)\alpha(2\tau,W(T))}.
\label{eqn:T2SpectralDiffusion}
\end{equation}
Here, $A_0(T)$ is the initial echo amplitude. The first exponential accounts for the intrinsic TLS A decoherence rate.
The second one expresses the spectral diffusion effect, where $\alpha(2\tau,W(T))$ is a function that averages over all the possible flip histories of TLSs B. The parameter $W$ is the temperature dependent TLS B jump rate (see Appendix~\ref{supp:Spectral_Model}). We assume this rate to be set by a single-phonon relaxation process $W=\Gamma_1^{\mathrm{B}}  \coth(\hbar \omega_{\mathrm{B}}/(2k_\mathrm{B}T))$~\cite{black1977,burin2015}.

\begin{figure}[h]
	\includegraphics{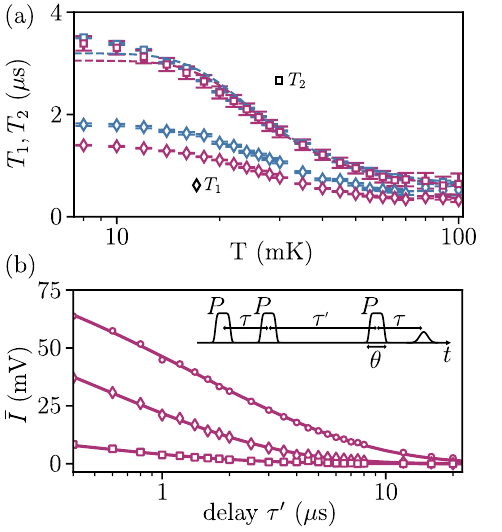}
    	\caption{(a) Extracted $T_1$ (diamonds) and $T_2$ (squares) versus temperature $T$ for JTWPA D2 (blue) and D3 (purple). Values for $T_2$ are obtained by a simple exponential fit while $T_1$ values are obtained from a stretched exponential fit. Dashed lines are predicted $T_2$ using the spectral diffusion model for device D2 (blue) and D3 (pink). (b) Mean echo amplitude $\bar{I}$ versus delay $\tau'$ for JTWPA D3 using a stimulated echo sequence (inset), with $\tau=\qty{0.55}{\micro\second}$ and $\theta = \qty{100}{\nano\second}$ for $\qty{8}{\milli\kelvin}$ (circles), $\qty{40}{\milli\kelvin}$  (diamonds) and $\qty{80}{\milli\kelvin}$  (squares). Purple lines: stretched exponential fits. 
        }
    \label{fig.3}
\end{figure}

\begin{table}[]
\begin{tabular}{|c|c|c|c|c|}
\hline
\multirow{ 2}{*}{JTWPA} & $\Gamma_2/(2\pi)$& $\Gamma_{\mathrm{sd}}^{0}/(2\pi)$& $\Gamma_1^{\mathrm{B}}/(2\pi)$ & $\omega_{\mathrm{B}}/(2\pi)$\\

 & (kHz) &  (kHz)& (kHz)  &(GHz) \\ \hline
D2     & $\num{50\pm 2}$& $\num{743\pm 87}$& $\num{146\pm 19}$& $\num{1.9\pm 0.1}$         \\ \hline
D3     & $\num{52\pm 2}$& $\num{831\pm 76}$& $\num{165\pm 17}$& $\num{2.0\pm 0.1}$\\ \hline
\end{tabular}
 \caption{Fit results for the spectral diffusion model parameters discussed in the main text.}
 \label{table1}
\end{table}
To quantify this spectral diffusion effect, we measure Hahn-echo decays for temperatures ranging from $\qty{8}{\milli\kelvin}$ to $\qty{110}{\milli\kelvin}$ above which the signal is too weak to be measured. 
 We perform a global fit over all temperatures traces we have measured by taking $\Gamma_{\mathrm{sd}}^0$, $\omega_\mathrm{B}$, $\Gamma_1^{\mathrm{B}}$  and $\Gamma_2$ as global free fit parameters (see Appendix \ref{supp:FittingEchoes}). The resulting fit for various amplifiers and temperatures is shown in Fig.~\ref{fig.2}a and  Fig.~\ref{fig.2}b (see also Appendix~\ref{supp:FittingEchoes}), with the fit values given in Table~\ref{table1}. It successfully captures the decaying echo signal, with a marginal difference compared to a single-exponential fit. This is explained by the fact that we are in a regime where the spectral diffusion is neither slow ($W\tau\ll1$ where $\tau^2$ dependence would be expected) nor fast ($W\tau\gg1$ where $\sqrt{\tau}$ dependence would be expected)\cite{black1977}. Comparing the coherence time $T_2$ extracted using a simple exponential fit (squares in Fig.~\ref{fig.3}a) to what is expected from our spectral diffusion model (dashed lines, with $T_2$ defined as $A(T_2,T)/A_0 = 1/e$), we find a rather good match for their temperature dependence. The spectral diffusion model gives a precise indication about the effective frequency of bath B $\omega_{\mathrm{B}}/(2\pi)= \qty{2.0\pm0.2}{\GHz}$.

\section{Observation of three-pulse echoes}

We now investigate the longitudinal relaxation time $T_1$ associated to these defects, see Fig.~\ref{fig.3}b. We use a three pulse sequence as for stimulated Hahn echo experiments~\cite{schweiger2001}. It consists of three identical pulses at frequency $\omega_{\mathrm{d}}/(2\pi)=\qty{7}{\giga\hertz}$ with power of $\qty{-77}{\dBm}$($\qty{-81}{\dBm}$) at the JWTPA D2 (D3) input separated by delay $\tau=\qty{0.55}{\micro\second}$ and $\tau'$ (see Fig. \ref{fig.3}b). 
The first pulse creates an initial transverse coherence which is partially converted to a polarization by the second pulse. The polarization decays through $T_1$ processes during $\tau'$, and is converted back to a transverse coherence by the third pulse. Waiting an additional time $\tau$ enables to refocus the dipoles and to generate an echo whose amplitude is governed by both waiting times. Sweeping the delay $\tau'$ between the second and third pulse monitors the polarization decay through the echo amplitude. Observed decays at $\qty{8}{\milli\kelvin}$, $\qty{40}{\milli\kelvin}$ and $\qty{80}{\milli\kelvin}$ are shown in Fig.~\ref{fig.3}b. These decays are no longer simply exponential, but fit well to a stretched exponential $A \exp(-(t/T_1)^p)$ with  $p_{\mathrm{D}2}=\num{0.606(5)}$ and $p_{\mathrm{D}3}=\num{0.547\pm0.006}$. This is characteristic of dynamics governed by spectral diffusion~\cite{black1977} (See Appendix~\ref{supp:Spectral_diffusion_Wex_model}). The extracted values of $T_1$ show a similar trend in temperature as $T_2$, with a decrease of $T_1$ around $\qty{30}{\milli\kelvin}$ (see Fig.~\ref{fig.3}a). 

\section{Impact on the amplifier performance}

The presence of these microscopic dielectric defects represents a source of dissipation for the JTWPA that ultimately degrades the signal to noise ratio at its output, and namely its quantum efficiency $\eta$~\cite{caves1982,macklin2015,houde2019,yuan2022}. Indeed, assuming a distributed loss model and that amplification is constant across all JTWPA cells, one finds $\eta =\frac{ag-1}{g - 1}$, where $a$ and $g$ are respectively the absorption and gain per JTWPA cell~\cite{macklin2015}. This efficiency is only an upper bound, since there exists others limiting effects such as parasitic reflections \cite{PhysRevB.107.174520} and hot environments \cite{PhysRevApplied.17.044009}. 
Assuming all losses per cell are due solely to defects in the dielectric capacitor, one can express $a$ as a function of $\tan \delta$ (see Appendix~\ref{supp:TWPAmodel}). The dielectric losses at zero temperature can be related to the concentration of TLS through $\tan \delta = \frac{4 \pi^2}{3 \epsilon} N_0 d_{\mathrm{A}}^2$ \cite{vonschickfus1977,burin2013}, where $N_0$ is the number of TLSs per unit of volume and per unit of energy. 

We now assess whether these dielectric echoes offer a way to calibrate in-situ the dielectric losses, namely whether we can evaluate finely $N_0$ and $d_{\mathrm{A}}$. One first possibility is to use the spectral diffusion rate since the underlying effect is the dipolar interaction between TLSs of bath A and of bath B (see Eq.~\ref{eqn.gammaSD}). We can compute $\tan \delta$ from $\Gamma_{\mathrm{sd}}$ at the price of three approximations. The first two are commonly-made: we can assume the TLS distribution follows the universal law $N(\Delta,\Delta_0) = \frac{N_0}{\Delta_0}$\cite{black1977,Burin_2013_tand} so that we can consider TLSs A and B to have the same concentration. Second, we can reasonably consider that the mean dipoles of TLSs A and TLSs B are of equal strength~\cite{vonschickfus1977,Burin_2013_tand}. Finally, a third hypothesis on the spectral width $\gamma_{\mathrm{B}}$ of the TLSs B contributing to spectral diffusion is needed to relate $N_0$ to $c_{\mathrm{B}}$  through $N_0 = c_{\mathrm{B}}/(\hbar\gamma_{\mathrm{B}})$. Given that we expect TLSs to be everywhere, and only TLSs of non-zero polarization can contribute to spectral diffusion, we can roughly estimate that the spectral width of the TLSs B bath is set by their thermal energy so that $\gamma_{\mathrm{B}}\sim\omega_{\mathrm{B}}$.
We thus find $\tan \delta = \frac{6\sqrt{3}\pi\Gamma_{\mathrm{sd}}^0}{\omega_{\mathrm{B}}}=\num{0.012\pm0.001}$ ($\num{0.014\pm0.001}$), and $\eta = \num{0.59\pm0.04}$ ($\eta = \num{0.54\pm0.04}$) for JTWPA D2 (D3). Our estimation of the loss tangent is about three times higher than previous reported values~\cite{macklin2015}. Let us note that the uncertainty on $\tan\delta$ comes from the fit of our spectral diffusion model and are highly optimistic ($\num{1e-3}$) compared to the level of confidence of our assumptions. A more in-depth study of the spectral diffusion mechanism and its microscopic origin is however necessary to refine these assumptions (see Appendix~\ref{supp:Spectral_diffusion_Wex_model}).

Another approach is to quantify $N_0$ and $d_{\mathrm{A}}$ using directly the two-pulse echo signal. To estimate these quantities, we model the bath of TLSs $A$ as an ensemble of TLSs coupled to a transmission line, where each TLS has a radiative coupling  to the line $\Gamma_{\mathrm{R}}$. When applying a pulse of amplitude $|\alpha_{\mathrm{in}}| = \sqrt{\frac{P_{\mathrm{in}}}{\hbar\omega_{\mathrm{d}}}}$ and of length $\theta$, a resonant TLS undergoes Rabi rotations at rate $\Omega=2\sqrt{\Gamma_{\mathrm{R}}}|\alpha_{\mathrm{in}}|$.  Using our experimental observation of Rabi oscillations in Fig.~\ref{fig.2}a, we can thus determine the value of $\Gamma_\mathrm{R}$. This Rabi oscillation can also be expressed through $\hbar \Omega = d_{\mathrm{A}} E_d$, where $E_d$ is the strength of the driving electrical field applied on the capacitor. Using this relation, we find $d_{\mathrm{A}}= \num{3\pm1}$D. This value is compatible with what is expected for $\text{OH}^-$  impurities \cite{baier1988,SiO2_dipole_strength} in silica. Next, we can estimate that the peak echo amplitude is proportional to the number of excited TLSs, so that $|\alpha_{\mathrm{out}}| =(N_0 V \gamma_{\mathrm{A}}) \sqrt{\Gamma_{\mathrm{R}}}$ where $V=\qty{1.5e-13}{\meter^{3}}$ is the volume of the capacitors comprising the JTWPA chain~\cite{macklin2015}, and $\gamma_{\mathrm{A}} = 2\pi/\theta$ is the bandwidth of excited TLSs, given by the duration of the first pulse. The precision of this technique suffers from our lack of precise power calibration and from non-canonical Rabi oscillations.  We find $N_0=\num{3\pm2e43}~\unit{\joule^{-1}\meter^{-3}}$, which lies close to values reported in the literature for amorphous materials~\cite{gao2008,Burin_2013_tand}. In this case, we find $\tan \delta = \num{18\pm16e-4} $ and  $\eta = \num{0.94\pm0.05} $ for JTWPA D1. Here the uncertainties are governed by the quality of the calibration of our lines. Despite the strong assumptions in both techniques, the results obtained indicate that, with additional characterization and modeling (see Appendix~\ref{supp:Spectral_diffusion_Wex_model}), these dielectric echoes could become an in-situ technique for characterizing dielectric losses in a JTWPA.

\section{Preventing the generation of dielectric echoes.}
\begin{figure}[h]
\includegraphics{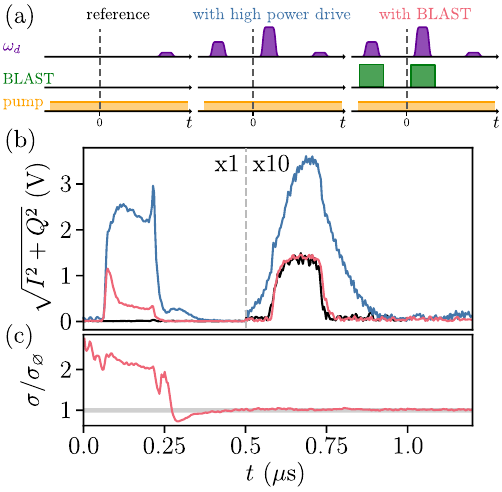}
\caption{Suppression of an echo by BLAST pulses, enabling to detect a small signal otherwise hidden. Sequences are shown in panel (a) and are run with the pump of the JTWPA $\mathrm{D}1$ always on. The amplitude of the detected outgoing signals at $\omega_{\mathrm{d}}$ are shown in panel (b). Reference (black): only a test signal at $\omega_{\mathrm{d}}/(2\pi) = \qty{7}{\GHz}$ of power $P_{\mathrm{probe}}=\qty{-125}{\dBm}$ referred to JTWPA input is applied and detected. With high-power drive (blue): two pulses at $\omega_{\mathrm{d}}$ ($P_1=\qty{-85}{\dBm}$, $P_2=\qty{-79}{\dBm}$, $\theta=\qty{150}{\nano\second}$) are applied before the probe signal with a timing such that the dielectric echo arises while the probe pulse is applied, resulting in observing the addition of the echo with the reference signal. With BLAST (red): square pulses are applied concurrently with $P_1$ and $P_2$ at $\omega_{\mathrm{BLAST}}/(2\pi)=\qty{6.14}{\giga\hertz}$ and power $P_{\mathrm{BLAST}}=\qty{-55}{\dBm}$, resulting in recovering the reference signal: the echo is suppressed and the JTWPA gain is preserved. (c) Ratio between the signal standard deviation $\sigma$ when performing BLAST and when no input pulses are applied $\sigma_{\varnothing}$. The standard deviation is calculated using $\sigma = \sqrt{\sigma^2_\mathrm{I} + \sigma^2_\mathrm{Q}}$, where $\sigma_{\mathrm{X}}(t)$ is the standard deviation of 5000 voltages recorded on quadrature $X$ at time $t$: $\sigma_\mathrm{X}^2=\langle X(t)^2-\bar{X}^2(t)\rangle$. Gray shadow: normalized reference standard deviation when no pulses are applied taking into account fluctuations from measurement to measurement due to room temperature setup drifts.}

\label{fig.4}
\end{figure}

We now present solutions for avoiding these dielectric echoes in experiments where they would mask a signal to be detected by the JTWPA. Possible situations are when high-power pulses are used to trigger the emission of a signal from a DUT at the same frequency of the drive pulses but at a later time, such as performing gates on a fluorescent qubit, or observing spin echoes in hybrid superconducting circuits. In the simplest conceivable setup, these high-power pulses are transmitted alongside the signal to be detected to the JTWPA. However, they will also trigger the emission of a dielectric echo in the JTWPA, which will mask the detection of the signal emitted by the DUT. Our global mitigation strategy thus relies on preventing the high-power pulses from reaching the JTWPA. One possibility is to cancel them interferometrically~\cite{eichler2012} in between the JTWPA and the DUT. While feasible (see Appendix~\ref{supp:interferometry}), it finely depends on all the components in the microwave setup and requires constant recalibration to account for phase drifts. 

We propose a second, more resilient protocol which can be complementary to interferometric pulses. It consists in using a microwave tone of far greater power ($+\qty{25}{\dB}$) than our high-power pulses (from $\num{-85}$ to $\qty{-75}{\dBm}$), or JTWPA pump ($\qty{-82}{\dBm}$) at a different frequency than either the drive/signal or pump tone. Such a high-power BLAST pulse turns the JTWPA into a reflective device (see Appendix \ref{supp:ReflectiveJTWPA}). Adding such a BLAST tone while we are sending high-power pulses reduces significantly the amount of power reaching the JTWPA, and thus should minimize the spurious dielectric echo generation.

We test the effectiveness of these BLAST pulses at canceling the unwanted echo and assess whether they permit to detect a dummy signal while achieving routine performance for the JTWPA. In addition to the two-pulse sequence used to generate an echo in Fig.~1b, we send a test signal at the same frequency $\omega_{\mathrm{d}}$ at the time at which the dielectric echo is occurring. Its power ($\qty{-125}{\dBm}$) is chosen to be much weaker than the JTWPA saturation threshold. Without using BLAST pulses in conjunction to high-power pulses, we observe the parasitic dielectric echo on top of this  signal (blue curve in Fig.~\ref{fig.4}b). When we add the BLAST pulses to the sequence (red), we recover the signal expected when sending only the test signal (black). We realize this experiment in presence of the JTWPA amplification pump tone, demonstrating that these BLAST pulses do not degrade the JTWPA gain and remove entirely the dielectric echo. We also measure the noise throughout the sequence, detecting no additional noise after a $\qty{300}{\nano\second}$ recovery time consecutive to the last BLAST pulse. Using BLAST pulses, we thus show that the JTWPA performs identically in terms of noise figure during the dummy signal measurement time without a parasitic echo. Let us note that the precise frequency of the BLAST pulse does not require a very fine calibration for this shielding effect to occur, and its power needs only be sufficient to make the JTWPA reflective (see Appendix \ref{supp:BLAST}).

\section{Conclusion}

In short, our results evidence the presence of dielectric echoes in the JTWPA due to microscopic defects in the JTWPA dielectric layers. These echoes need to be taken into account in many experiments, but can be simply avoided through the application of a BLAST pulse. By analyzing the strength of these echoes signals and their time dynamics using a spectral diffusion model, we can also estimate the internal losses they induce in the JTWPA and the limit they impose on quantum efficiency, leading the way for these measurements to become an additional in-situ characterization technique for JTWPAs. Future experiments could probe traveling-wave amplifiers beyond the $4$ to $\qty{8}{\giga\hertz}$ range studied here to compare with other experimental results~\cite{ProbeTLS1,ProbeTLS2}. It would allow to access for the first time the frequency distribution of these microscopic defects. In addition, it would be interesting to check whether it is possible to correlate these dynamics to effects on other quantities such as the intermodulation product, or to material changes in JTWPA devices.

\section{Data availability}
The data supporting the findings of this study are available in Zenodo at https://doi.org/10.5281/zenodo.16900112

\begin{acknowledgments}
This work was supported by the European Union (ERC, INDIGO, 101039953) and by the ARO GASP (contract No. W911-NF23-10093) program. We acknowledge IARPA and Lincoln Labs for providing us with Josephson Traveling-Wave Parametric Amplifiers. We thank William Oliver, Alexis Coissard, Romain Albert, Luca Planat and Nicolas Roch for fruitful discussions.
\end{acknowledgments}
\bibliography{twpa}

\newpage
\onecolumngrid
\newpage

\appendix
\makeatletter 
\renewcommand{\thefigure}{A\@arabic\c@figure}
\makeatother

\section{Experimental setup}\label{supp:fridge_setup}
The JTWPA is driven by pulses generated with a Zynq UltraScale+ RFSoC ZCU216 Evaluation Kit, set up with QICK firmware~\cite{qick}. The driving pulses are sampled at $\qty{430.08}{\mega\hertz}$ and modulated at a frequency $\qty{193}{\mega\hertz}$. They are up-converted using I-Q mixers, with continuous microwave tones produced by one of the four 
channels of an AnaPico® APUASYN20-4 generator, and then are amplified by a Mini-Circuits ZVE-3W-183+ amplifier followed by a Wainwright Instruments WTBCX6-6500-7000-40-200-40SS bandpass filter to reject the local oscillator leakage before reaching the fridge input. 
A second channel of the generator is used for the JTWPA pump, while a third channel is dedicated to the BLAST tone. To characterize the JTWPA D1 amplification in Fig.~4, we tune the JTWPA pump frequency ($\omega_{\mathrm{p}}/(2\pi)=\qty{5.985}{\giga\hertz}$) in order to reach a sweet point of gain $G=\qty{20.8(1)}{\dB}$ and  noise figure $F=\qty{13.2(1)}{\dB}$ for the entire amplification chain when the JTWPA is turned on and off. The BLAST tone output is controlled via an RF-Lambda Absorptive Coaxial SPST Switch 2GHz-18GHz.
The echoes emitted by the JTWPAs are amplified by a HEMT amplifier from Low Noise
Factory® at 4 K and by a room-temperature amplifier. The signals are down-converted using I-Q mixers before digitization by the ZCU216 board with readout sampling readout rate of $\qty{307.2}{\mega\hertz}$. 
\begin{figure*}[h!]
\includegraphics{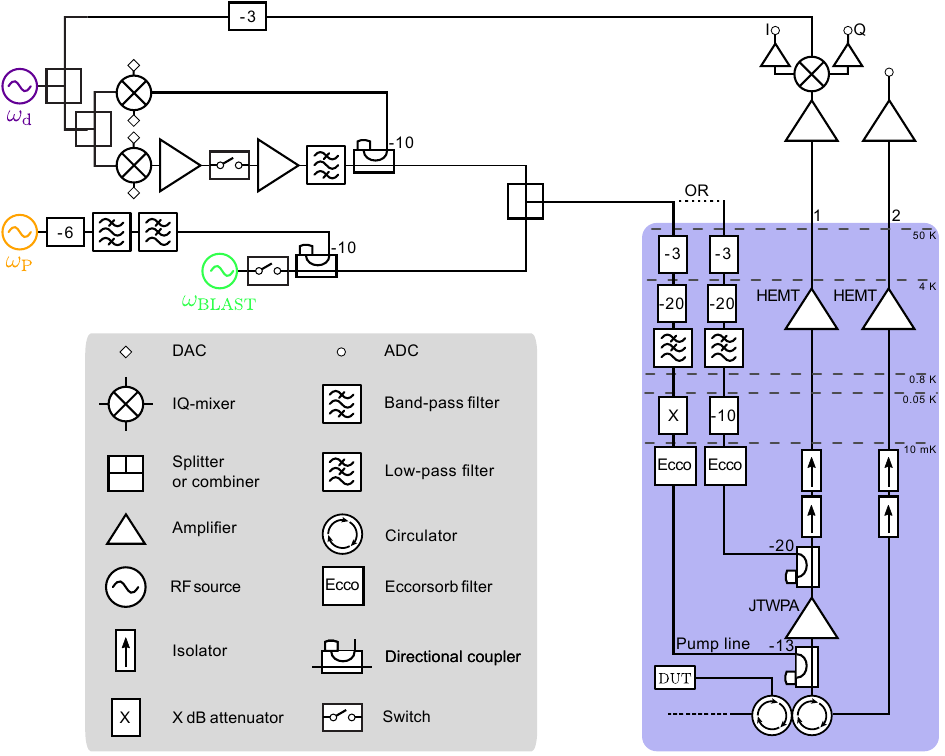}
\caption{General schematic of the measurement setup for JTWPA D1, D2 and D3. In the measurements presented in the main text, we send our drive signals directly through the pump line. For JTWPA D1, we placed a $\qty{-10}{\decibel}$  attenuator on the cold plate (0.05K), and we use a $\qty{-13}{\decibel}$ directional coupler to combine the signal from the DUT and the JTWPA pump. JTWPA D2 and D3 were measured in another fridge of nominally identical wiring, except the cold plate attenuator was $\qty{-20}{\decibel}$ and the directional coupler had $\qty{-20}{\decibel}$ coupling.  When measuring  JTWPA D2 and D3, we omitted the directional coupler placed directly after the JTWPA and the DUT is replaced by a $\qty{50}{\ohm}$ termination. On top of the fridge, we can choose whether to probe the JTWPA or drive directly the HEMT. We do not represent room-temperature isolators for clarity. }

\label{fig.fridge_setup}
\end{figure*}

\section{Echo signal integration}\label{supp:fitEchoRoutine}

\begin{figure*}[h!]
\includegraphics{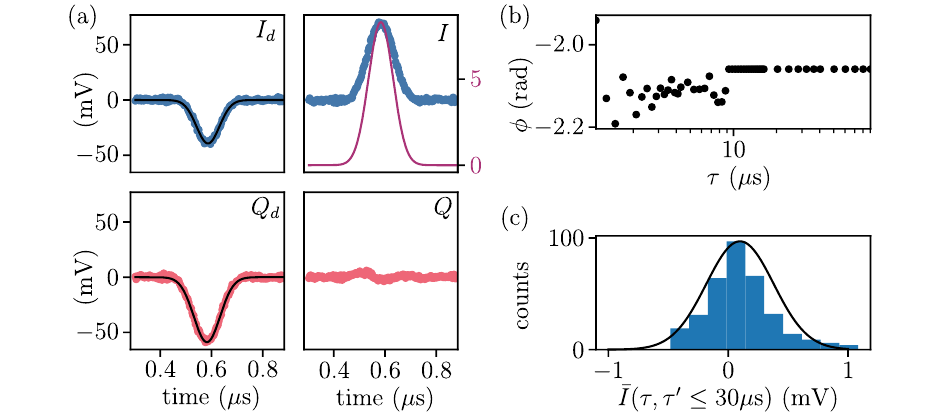}
\caption{(a) Time traces of the detected $I_{\mathrm{d}}$ (blue) and $Q_{\mathrm{d}}$ (red) quadrature taken on JTWPA D2 at $\qty{8}{\milli\kelvin}$ and $\tau=\qty{0.6}{\micro\second}$, 
taken with JTWPA pump off, $P_1=\qty{-81}{\dBm}$, $P_2=\qty{-75}{\dBm}$, and $\omega_\mathrm{d}/(2\pi)=\qty{7.0}{\giga\hertz}$.
After rotating by an angle $\phi$, we obtain the $I$ and $Q$ quadratures. Black, Gaussian fit on $I_{\mathrm{d}}$ and $Q_{\mathrm{d}}$. Purple, filtering function $u(t)$. (b) $\phi$ as function of $\tau$ for JTWPA D2 at $\qty{8}{\milli\kelvin}$ (c) Histogram of the signals used to evaluate the error on $\bar{I}$. Black, Gaussian fit of the point distribution. }
\label{fig.fitting_echo}
\end{figure*}

We compute the mean average signal $\bar{I}$ as the weighted integrated amplitude: $\bar{I}=\int u(t) I(t)dt / \int u(t)^2 dt$, where $u(t)$ is a Gaussian filter function having the same center time and width as the spurious signal

We now detail the procedure to extract the weighted integrated amplitude for each echo signal. We first detect the $I_d$ and $Q_d$ quadrature using heterodyne detection at $\omega_\mathrm{d}$. The sampling time is $\Delta t = \qty{3.2}{\nano\second}$. We then fit every echo signal $j$ with a Gaussian $Ae^{\frac{-(t - \mu)^2}{2\sigma ^2}}$. We independently treat the $I_{\mathrm{d}}^j$ and $Q_{\mathrm{d}}^j$  quadratures, thus obtaining $\mu_I^j$,$\mu_Q^j$,$\sigma_I^j$,$\sigma_Q^j$. 
In a spontaneous or stimulated Hahn echo sequence, we use the three highest echo signals i.e. the echoes detected for the three shortest delays, to define a Gaussian filtering function $u(t) = \frac{1}{\sqrt{2\pi\bar{\sigma}^2}}e^{\frac{-(t-\bar{\mu})^2}{2\bar\sigma^2}}$ where $\bar\mu = \frac{1}{6}\sum_{j=1,2,3}(\mu_{I}^j+\mu_{Q}^j)$ and $\bar\sigma = \frac{1}{6}\sum_{j=1,2,3}(\sigma_{I}^j+\sigma_{Q}^j)$. We also define in the complex IQ plane the angle $\phi_0 =\angle \biggl\{ \sum_{j=1,2,3} \left[ \sum_{i=1} ^N I_{\mathrm{d}}^j(T^i)+i\sum_{i=1} ^N Q_{\mathrm{d}}^j(T^i) \right] \biggl\}$. Correcting for the phase rotation, we can access to the 
weighted integrated amplitude by computing:
\begin{equation}
\bar{I}=\frac{\int_{t_0}^{t_1} u(t) I(t)dt}{\int_{t_0}^{t_1} u(t)^2 dt}=\sum_{i=1}^{N} u(T^i) \left[ I_{\mathrm{d}}(T^i)\cos(\phi)+Q_{\mathrm{d}}(T^i)\sin(\phi) \right]\Delta t.
\end{equation}
Here $u(t)$ is a Gaussian filter function having the same center time and width as the spurious signal and $\phi=\phi_0 + \delta$ where $\delta$ is a small correction that minimizes $\bar{Q}=\sum_{i=1}^{N} u(T^i) \left[ -I_{\mathrm{d}}(T^i)\sin(\phi)+Q_{\mathrm{d}}(T^i)\cos(\phi) \right]\Delta t$ for each echo. An example of the optimal $\phi$ obtained for each $\tau$ is shown in Fig.~\ref{fig.fitting_echo}b. 

To evaluate the error on $\bar{I}$, we collect all measurements performed using the same room temperature setup with $\tau>\qty{30}{\micro\second}$ and $\tau'>\qty{30}{\micro\second}$ for spontaneous and stimulated Hahn echo experiments. Since no echo is present in such traces, what we measure is only the noise weighted by $\mu(t)$. Fitting the histogram of such points with a Gaussian (see Fig.~\ref{fig.fitting_echo}c), we extract $\sigma(\bar{I})=\qty{0.28}{\milli\volt}$.

\newpage
\section{Spectral diffusion modeling}
\label{supp:Spectral_Model}
\subsection{Sudden-jump model}
We discuss here the decoherence induced by spectral diffusion. Let us consider a defect subjected to static strain and/or electric fields which depend on the host material. A double well potential with asymmetry splitting $\Delta$ and tunneling energy $\Delta_0$ is widely used to model such system. The Hamiltonian is analogous to the one of an electronic spin 1/2 in a static magnetic field $\vec{B}_{\mathrm{0}}$: 
\begin{equation}
H_A/\hbar=\frac{\Delta}{2}\sigma_z + \frac{\Delta_0}{2} \sigma_x
\end{equation}
with $\sigma_x=
\begin{pmatrix}
0 & 1 \\
1 & 0
\end{pmatrix}$ and $\sigma_z=
\begin{pmatrix}
1 & 0 \\ 0 & -1
\end{pmatrix}$. 
These TLSs are coupled to one another elastically or electrically. We focus here on electric dipole-dipole interaction between TLS A with dipole $d_{\mathrm{A}}$ and a bath of B TLSs comprising $N_{\mathrm{B}}$ dipoles $d_{\mathrm{B}}^k$ distant from $A$ by $\vec{r}_k$ with $k=1,2,..N_{\mathrm{B}}$. Since the interaction is short-range we assume the local static electric field $\vec {E}_0$ to be homogeneous.
We consider only the case of non-resonant TLSs, for which 
\begin{equation}
H_{\mathrm{AB}}=\sum_{k=1} ^NC_{\mathrm{AB}}^k\sigma_z^{\mathrm{A}} \sigma_z^{\mathrm{k}}
\end{equation}
with $C_{\mathrm{AB}}^k=\frac{1-3\cos^2(\theta^k)}{4\pi\epsilon (r_k)^3 }d_{\mathrm{A}}d_{\mathrm{B}}^k$
where $\theta^k$ is the angle between $\vec{E}_0$  and $\vec{r}_k$.
The TLS A energy $E_a = \hbar \omega_{\mathrm{A}}$ thus depends on the state of TLS B bath as 
\begin{equation}
\omega_{\mathrm{A}}(t)=\omega_{\mathrm{A}}^0 + \sum _{k=1} ^N\frac{C_{\mathrm{AB}}^k}{\hbar}\sigma_{z}^{k}(t).
\end{equation}
TLSs B are thermally active and can jump between their ground and excited state having thus an impact on TLS A. This effect is known as spectral diffusion. Once TLS A is excited its angular frequency $\omega_{\mathrm{A}}(t=0)$ is modified at each TLS B jump, drifting away from its starting value.
We focus now on the effect of spectral diffusion in an Hahn echo sequence (see main text). The echo amplitude of TLS A is given by:
\begin{equation}
	E(2\tau) = \Re\left(e^{i\int _0 ^\tau \omega_{\mathrm{A}}(t)dt - i\int _\tau ^{2 \tau} \omega_{\mathrm{A}}(t)dt }\right)
\end{equation}
where the exponent can be rewritten as:
\begin{equation}
\sum _{k=1} ^N\frac{C_{\mathrm{AB}}^k}{\hbar}\int _0 ^{2\tau} s(t)h^k(t)dt
\end{equation}
where $h^k(t)$ is a function taking values $\pm1$ and changing sign at every jump of the $k$ TLS while $s(t)=
\begin{cases}
      1 & t\le \tau\\
      -1 & t>\tau
\end{cases}$.

The echo amplitude thus depends on the spatial distribution and on the flipping history of the B TLSs bath. In order to treat this problem we approximate the TLS bath as an ensemble of identical B TLSs of mean dipole $\bar{d}_{\mathrm{B}}$ where we average over all B TLSs positions and flip histories (FP) and consider a mean interaction strength $\bar{C}_{\mathrm{AB}}$. Following~\cite{hu1974}, we can express the echo amplitude as: 
\begin{equation}
E(2\tau) = A_0 \exp\left[ -\frac{2\pi  }{9\sqrt{3}\hbar\epsilon}d_{\mathrm{A}}\bar{d}_{\mathrm{B}} c_{\mathrm{B}}  \langle\left|\int _0 ^{2\tau} s(t)h(t)dt\right|\rangle_{\text{FH}}\right]
\end{equation}
where $c_{\mathrm{B}}$ is the spatial density of flipped B TLSs. The average over all possible B flip histories can be conducted considering a single flipping rate $W$. The derivation can also be found in \cite{hu1974}, obtaining
\begin{equation}
\langle\left|\int _0 ^{2\tau} s(t)h(t)dt\right|\rangle_{\text{FH}} =  \alpha(2\tau,) = 2e^{-2W\tau}\tau\left[I_1(2W\tau) +\frac{\pi}{2}\left( I_1(2W\tau)L_0(2W\tau) -I_0(2W\tau)L_1(2W\tau) \right)\right]
\end{equation}
where $I_i(x)$,$L_j(x)$ are the modified Bessel and Struve function of order $i,j$. Defining $\Gamma_{\mathrm{sd}} = \frac{2\pi  }{9\sqrt{3}\hbar\epsilon}d_{\mathrm{A}}\bar{d}_{\mathrm{B}} c_{\mathrm{B}}$ we obtain the expression $E(2\tau)$ of the main text.\\
For the three pulse sequence it is possible to proceed in a similar way:
\begin{equation}
    E(2\tau,\tau') =  \exp\left[ -\Gamma_{\mathrm{sd}} \langle\left|\int _0 ^{\tau} h(t)dt - \int_{\tau+\tau'} ^{2\tau+\tau'} h(t)dt\right|\rangle_{\text{FH}}\right]=\exp\left[ -\Gamma_{\mathrm{sd}} \beta(\tau,\tau',W)\right] 
\end{equation}
where there is no precession of the TLS A between the second and third pulse. Averaging over all B TLSs flipping histories we obtain
\begin{equation}
\beta(\tau,\tau',W) = e^{-2W\tau}\tau\left[I_0(2W\tau) + I_1(2W\tau)\right]\left(1-e^{-2W\tau'}\right) +\frac{1}{2}\alpha(2\tau,W)\left(1+e^{-2W\tau'}\right).
\end{equation}

\subsection{Fitting two-pulse echoes}\label{supp:FittingEchoes}

\begin{figure}[h]
    \centering
    \includegraphics{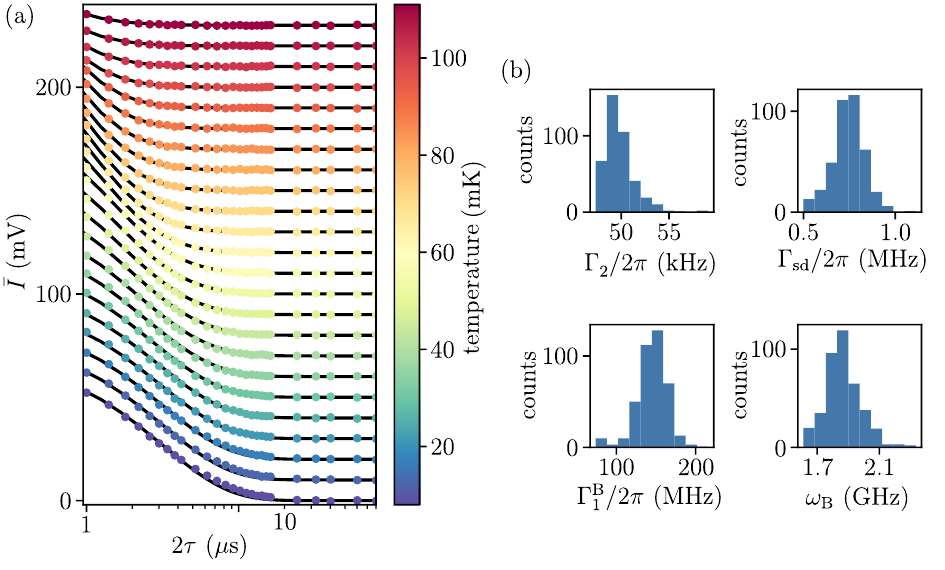}
    \caption{Modeling without including bath dynamics. (a) Two pulse sequence integrated echo versus delay $2\tau$ for all measured temperatures.  We show half of the experimental points and there are 10 mV offset between two consecutive temperatures for clarity. Black lines correspond to spectral diffusion model fit (see Eq.~\ref{eqn:T2SpectralDiffusion}). (b) Histograms of the optimal parameters in spectral diffusion model. All data presented refer to JTWPA D2.} 
    \label{fig:TWPA608_bootstrappNoB}
\end{figure}

We detail the routine implemented for fitting the two pulse experiments, considering as we have done in the main text that in addition to the spectral diffusion dynamics, the coherence of the echo is governed by a temperature-independent decoherence rate. There are 28 parameters in total namely $\Gamma_2,\Gamma_{\mathrm{SD}}^0,\Gamma_1^{\mathrm{B}},\omega_{\mathrm{B}}$ and the twenty-four amplitudes $A_0(T^i)$  $A_0(T^i)$ corresponding to the twenty-four temperatures $T^i$ probed. Due to the large parameter space, we use bootstrapping to evaluate all parameters and their variance. We focus on the two pulse experiment. We have acquired 24 echoes relaxation traces, each at a different temperature, with 52 integrated echoes per trace acquired for various $\tau$. We can randomly choose a subset of 18 of these traces to perform the fit and obtain a fit vector for our 28 parameters. Repeating the operation four hundred times, we obtain four hundred fit results, from which we can extract the mean and variance of each of these fit parameters. The fit of each data-subset is performed by minimizing the cost function
\begin{equation}
    C = \sum_{i=1}^{18} \left[\sum_{\tau}\left[\bar{I}(\tau,T^i) -A_0(T^i) e^{-2(\Gamma_2\tau} e^{-\Gamma_{\mathrm{sd}}^0(\omega_{\mathrm{B}},T^i)\alpha(2\tau,\Gamma_1^{\mathrm{B}}(\omega_{\mathrm{B}},T^i))}\right]^2\right]
\end{equation}
where $I(\tau,T^i)$ is the integrated echo with delay $\tau$ and temperature $T^i$, $A_0(T^i)$ is the amplitude of the echo at temperature $T^i$ and the temperature independent parameters discussed in Eqn.\ref{eqn:T2SpectralDiffusion} of the main text. We thus obtain a distribution for the global fit parameters, shown in Fig.\ref{fig:TWPA608_bootstrappNoB}(b) from which we extract the mean and standard deviation values presented in table \ref{table1}. The fitting curves are presented in Fig.\ref{fig:TWPA608_bootstrappNoB}(a).

\subsection{Fitting three-pulse echoes}

\begin{figure}[h!]
\includegraphics{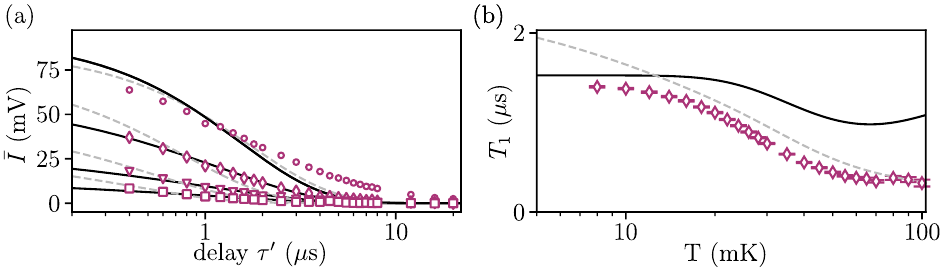}
\caption{Modeling $T_1$ dynamics. (a) Purple markers show the measured echo amplitude for the three-pulse echo sequence acquired at $T=\qty{8}{\milli\kelvin}$, $\qty{40}{\milli\kelvin}$, $\qty{60}{\milli\kelvin}$ and $\qty{80}{\milli\kelvin}$. The black lines represent the expected decay rate using a spectral diffusion model that includes a temperature independent intrinsic coherence rate (defined in the main text, see Eq.\ref{eqn:T2SpectralDiffusion}). The gray dashed lines correspond to a model assuming a temperature-dependent coherence rate (defined in Appendix.~\ref{supp:Spectral_diffusion_Wex_model}, see Eq.~\ref{eqn:T2_with_Wex}). (b) $T_1$ rates as extracted from the data and the two different models.}
\label{fig:SUPP_T1Models}
\end{figure}

We investigate whether the Hahn echo model can predict also the three pulses stimulated Hahn echo results we have obtained, see Fig. \ref{fig.3}(b). The same sudden jump model we used for Hahn echo decay (see Eqn. \ref{eqn:T2SpectralDiffusion}) can also model the stimulated echo amplitude:
\begin{equation}\label{eqn.SHE}
A^{\mathrm{se}}(\tau',T) = A_0^{\mathrm{se}}(T) e^{-\Gamma_1\tau'}e^{-\Gamma_{\mathrm{sd}}^0 \beta(\tau,\tau',W)}.
\end{equation}
Here, $A_0^{\mathrm{se}}$ is the initial amplitude. The second exponent captures the effect of spectral diffusion with $\beta$ a function taking into account the jump history similarly to $\alpha$~\cite{hu1974}, and fully determined by the fit of the Hahn echo signals. The first exponential describes an intrinsic energy relaxation $\Gamma_1$, and we make the assumption that it fully sets the intrinsic decoherence rate $\Gamma_1 = 2\Gamma_2$ extracted from fitting the two-pulse echoes shown in Fig.\ref{fig.3}(a). 
We fit our data using the parameters in table \ref{table1} adjusting $A_0^{\mathrm{se}}$ for each trace. The other parameters are taken from the fit of the two-pulses experimental data. In Fig.~\ref{fig:SUPP_T1Models}a, we show this fit for $T=\qty{8}{\milli\kelvin}$, $\qty{40}{\milli\kelvin}$, $\qty{60}{\milli\kelvin}$ and $\qty{80}{\milli\kelvin}$ (solid black lines). The model provides a good fit at high temperatures, but it degrades at lower temperature. Similarly to $T_2$, we can extract the $T_1$ values from our model using the relation $A^{\mathrm{se}}(T_1)/A_0^{\mathrm{se}}=1/e$. The results are plotted in Fig \ref{fig:SUPP_T1Models}, see solid black lines. The comparison of the full model with the simple stretch exponential rates indicates that at least another intrinsic relaxation mechanism plays a role. At temperatures $T\ll\hbar\omega_{\mathrm{B}}/\kappa_{\mathrm{B}}$ the B TLSs flipping rate is suppressed limiting thus the spectral diffusion effect. When looking at high temperatures $T$ such that $W\tau\gg1$, the B TLSs flip so fast that on average they do not affect the TLSs A frequency anymore, bringing no additional contribution to the decay rate $\Gamma_1$ \cite{hu1974}.

\newpage
\subsection{Refining the spectral diffusion model: taking into account intrinsic energy relaxation}\label{supp:Spectral_diffusion_Wex_model}

So far, we have assumed the TLSs bath dynamics is set by spectral diffusion associated only to an intrinsic temperature independent relaxation process, creating a temperature independent intrinsic decoherence. However, this simple explanation does not capture fully our dataset, namely it does not capture the relaxation process occurring at low temperatures, and it also overestimate the $T_1$ values we extract at higher temperatures. We investigate in this section a refined model that introduces an energy relaxation rate $\Gamma^{\mathrm{A}}_1$ of the excited TLSs A which is temperature dependent. 
There exist a few different mechanisms that can provide intrinsic energy relaxation beyond spectral diffusion for TLSs A. One could think of relaxation by single phonons,  which scales as  $\coth(\hbar \omega_{\mathrm{A}} / 2\kappa_{\mathrm{B}}T)$, but it is expected to be negligible for temperatures $T\ll \hbar \omega_{\mathrm{A}}/\kappa_{\mathrm{B}}$. What can be considered instead is a TLS-TLS interaction term that scales linearly with the temperature and becomes predominant at $T\ll \hbar \omega_{\mathrm{A}}/\kappa_{\mathrm{B}}$\cite{burin2013,ExchangeTerm}. We thus model $\Gamma_1=W_{\mathrm{ex}}T + \Gamma_1^*$, where $\Gamma_1^*$ is a temperature-independent relaxation rate.  We then keep the assumption that the intrinsic decoherence rate $\Gamma_2$ is purely limited by the energy relaxation $\Gamma_2(T) = \Gamma_1/2 = W_{\mathrm{ex}}T/2 + \Gamma_2 ^*$ with $\Gamma_2 ^*=\Gamma_1^*/2$.


\subsubsection{Fitting two-pulse echoes}
\begin{figure}[h]
    \centering
    \includegraphics{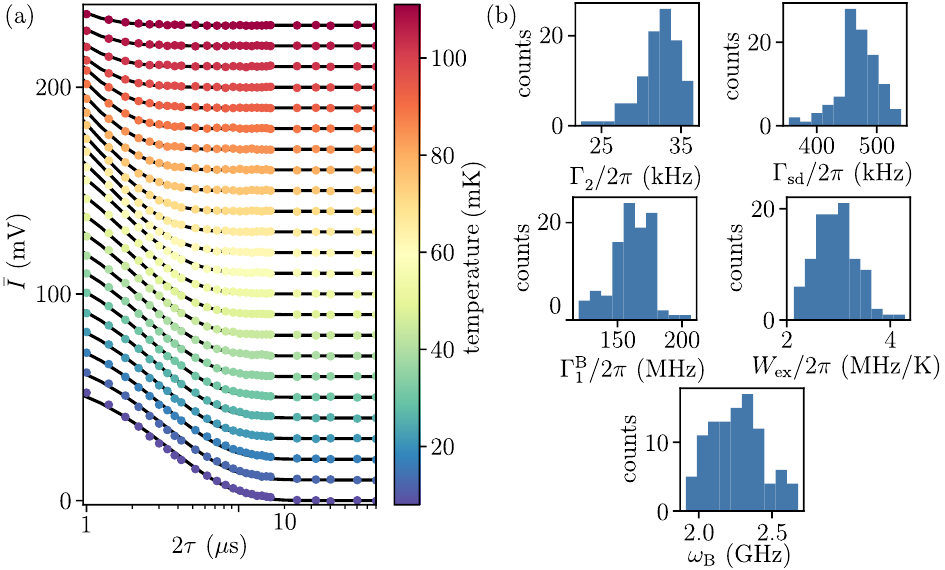}
    \caption{Modeling including bath dynamics. (a) Two pulse sequence integrated echo versus delay $2\tau$ for all measured temperatures.  We show half of the experimental points and there are 10 mV offset between two consecutive temperatures for clarity. Black lines correspond to spectral diffusion model fit (see Eq.~\ref{eqn:T2_with_Wex}). (b) Histograms of the optimal parameters in spectral diffusion model. All data presented refer to JTWPA D2.} 
    \label{fig:evaluate_fitallT2_Wex}
\end{figure}

Taking into account this new assumption, the Hahn echo amplitude versus temperature $T$ and delay $\tau$ now reads (see Eqn.\ref{eqn:T2SpectralDiffusion}):
\begin{equation}
A(2\tau,T)=A_0(T) e^{-2(W_{\mathrm{ex}}T/2 + \Gamma_2 ^*)\tau} e^{-\Gamma_{\mathrm{sd}}\alpha(2\tau,W)}.
\label{eqn:T2_with_Wex}
\end{equation}
We implement the same fitting routine as in Appendix \ref{supp:FittingEchoes}. There are now 29 parameters in total namely $\Gamma_2^*, W_{\mathrm{ex}},\Gamma_{\mathrm{sd}}^0,\Gamma_1^{\mathrm{B}},\omega_{\mathrm{B}}$ and the amplitude $A_0(T^i)$ with $T^i$ being the temperatures at which we performed the two-pulse echo measurement. We use bootstrapping, choosing randomly one-hundred subsets with 18 out of 24 the echoes relaxation traces, as done in Appendix \ref{supp:FittingEchoes}. The fit of each data-subset is performed by minimizing the cost function
\begin{equation}
    C = \sum_{T^i=1}^{18} \left[\sum_{\tau}\left[\bar{I}(\tau,T^i) -A_0(T^i) e^{-2(\Gamma_2^*+W_{\mathrm{ex}}T^i/2)\tau} e^{-\Gamma_{\mathrm{sd}}^0(\omega_{\mathrm{B}},T^i)\alpha(2\tau,W(\omega_{\mathrm{B}},T^i))}\right]^2\right]
\end{equation}
where $I(\tau,T^i)$ is the integrated echo with delay $\tau$ and temperature $T^i$, $A_0(T^i)$ is the amplitude of the echo at temperature $T^i$ and the temperature independent parameters discussed in Eq.~\ref{eqn:T2SpectralDiffusion} of the main text. We thus obtain a distribution for the global fit parameters, shown in Fig.\ref{fig:evaluate_fitallT2_Wex}(b) from which we extract the mean and standard deviation values presented in Table \ref{table:Wex}.
We then use the mean values of the temperature independent parameters together with Eq.~\ref{eqn:T2SpectralDiffusion} to obtain $A_0(T^i)$ for each temperature. The fitting curves are presented in Fig.\ref{fig:evaluate_fitallT2_Wex}(a) for JTWPA D2. We repeat the same fitting routine on device D3.

\begin{table}[h!]
\begin{tabular}{|c|c|c|c|c|c|}
\hline
\multirow{ 2}{*}{JTWPA} & $\Gamma_2^*/(2\pi)$& $\Gamma_{\mathrm{sd}}^0/(2\pi)$& $\Gamma_1^{\mathrm{B}}/(2\pi)$ &$W_{\mathrm{ex}}/(2\pi)$& $\omega_{\mathrm{B}}/(2\pi)$\\

 & (kHz) &  (kHz)& (kHz)  &(MHz/K)&(GHz) \\ \hline
D2     & $\num{32\pm 3}$& $\num{468\pm 34}$& $\num{161\pm 16}$&$\num{2.9\pm0.4}$& $\num{2.3\pm 0.2}$         \\ \hline
D3     & $\num{33\pm 2}$& $\num{586\pm 46}$& $\num{187\pm 19}$&$\num{3.0\pm0.4}$& $\num{2.4\pm 0.2}$\\ \hline
\end{tabular}
 \caption{Fit results for the spectral diffusion model discussed in the appendix~\ref{supp:Spectral_diffusion_Wex_model}, using Eq.~\ref{eqn:T2_with_Wex}.}
 \label{table:Wex}
\end{table}

\subsubsection{Fitting three-pulse echoes}
Using the same sudden jump model as before, we integrate our intrinsic energy decay mechanism for modelling the  three pulses stimulated Hahn echo results. The model writes as:
\begin{equation}\label{eqn.SHE_Exchange}
A^{\mathrm{se}}(\tau',T) = A_0^{\mathrm{se}}(T) e^{-\Gamma_1(T)\tau'}e^{-\Gamma_{\mathrm{sd}}^0 \beta(\tau,\tau',W)}.
\end{equation}
Here, $A_0^{\mathrm{se}}$ is the initial amplitude and the first exponential captures the intrinsic energy decay, where we assume $\Gamma_1(T) = \Gamma_1^*+W_{\mathrm{ex}}T$. The second exponential accounts solely for the effect of spectral diffusion and is again fully determined by the fit of the Hahn echo signals. 


We fit our data using the parameters in table \ref{table:Wex} adjusting $A_0^{\mathrm{se}}$ for each trace. The other parameters are taken from the fit of the two-pulses experimental data. In Fig.~\ref{fig.3}b, we show this fit for $T=\qty{8}{\milli\kelvin}$, $\qty{40}{\milli\kelvin}$, $\qty{60}{\milli\kelvin}$ and $\qty{80}{\milli\kelvin}$ (gray dashed lines). We find that the model is slightly better at capture the $T_1$ behavior at higher temperature, but also overestimates the relaxation at the lowest temperature. 



\subsubsection{Comparison between the two models}

Comparing the results with the model of the main text (see Table \ref{table1}), we remark that the spectral diffusion contribution $\Gamma_{\mathrm{sd}}$ is reduced about 1.5 times due to an increasing contribution of the first exponent in Eq.~\ref{eqn:T2_with_Wex}, highlighting the interest in better understanding the decoherence and relaxation mechanisms at play beyond spectral diffusion to better quantify the latter. 

\begin{figure}[h!]
    \centering
    \includegraphics{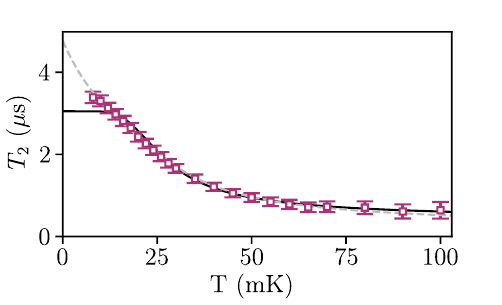}
    \caption{Comparison of the two sprectral-diffusion models to the measured JTWPA echo measured coherence, acquired for JTWPA D3. Purple markers represent the coherence rate extracted from simple exponential fits on the two-pulse echoes measurements realized as a function of temperature $T$, see main text. The black line represents the expected coherence time using a spectral diffusion model that includes a temperature independent intrinsic coherence rate (defined in the main text, see Eq.\ref{eqn:T2SpectralDiffusion}). The gray dashed line corresponds to a model assuming a temperature-dependent coherence rate (defined in Appendix.~\ref{supp:Spectral_diffusion_Wex_model}, see Eq.~\ref{eqn:T2_with_Wex}).}
    \label{fig:compareModelsT2}
\end{figure}

The extracted $T_2$ decoherence rates versus temperature using temperature independent $T_2$ (black line, model described in the main text) and temperature-dependent $\Gamma_2(T)$ (gray-dashed line, described in \ref{supp:Spectral_diffusion_Wex_model}) are shown in Fig. \ref{fig:compareModelsT2}. We see that there is no qualitative difference in the two models for capturing the decoherence processes, except for their behavior at low temperature, where the simpler model presented in the main text predicts a fixed coherence rate, and the model including bath dynamics predicts an infinite decoherence rate. Let us note that the latter behavior would be more aligned to what has been measured so far in the literature~\cite{bernard1979a}. Measuring the JTWPA echo dynamics either at lower temperatures, or at higher frequencies is then key in further refining our model to better capture the spectral diffusion process, and thus be better able to quantify the JTWPA dielectric losses. 

For the relaxation rate, it is harder to pinpoint whether one model significantly outperforms the other, as they each have their own weaknesses. Here again, measurements at higher frequencies and lower temperatures would be helpful. Other sequences than the three-pulse echoes could also be developed to better quantify the energy relaxation mechanisms independently of spectral diffusion. We have attempted implementing such sequences (such as broadband inversion or saturation recovery), but we found that the non-linearity of the JTWPA limited the amount of microwave power we could use to implement these sequences, which severely limits the performances of these sequences in measuring $T_1$ processes independently of spectral diffusion.

\newpage
\section{Dielectric losses impact on quantum efficiency}\label{supp:TWPAmodel}
We detail here how dielectric losses play a role in limiting the quantum efficiency of the JTWPA amplifier.  Focusing on a single cell of the JTWPA chain, we can model the dielectric losses in the capacitor $c$ as a conductance $\Upsilon$ in parallel to it:
\begin{equation}
\Upsilon=c \omega \tan\delta
\end{equation}
Such finite conductance acts as an attenuator between one cell and another. The attenuation can be expressed as 

\begin{equation}
a\approx
1-c\omega Z_0\tan\delta
\end{equation}
where $Z_0$ is the impedance of the JTWPA chain and we consider $c\omega Z_0\tan\delta\ll1$.\\
We can now consider the noise of $n_{\mathrm{c}}$ amplifying cells constituting the JTWPA. We model each cell as an attenuator of attenuation $a$ plus an ideal phase-preserving amplifier of gain $g$ and we assume all cells to be equal. We consider an input noise $N_0=N_{\mathrm{Q}}=\frac{1}{2}$ at the first cell input. To evaluate how the noise is modified going through the first cell, we need to account three terms, namely the cell attenuation $a$, the quantum fluctuation added by the the attenuator $(1-a)N_{\mathrm{Q}}$ and the noise $(1-1/g)N_{\mathrm{Q}}$ added by amplifier. The total noise at the output of the first cell is \cite{macklin2015}:
\begin{equation}
    N_1 = gaN_0 + g(1-a)N_{\mathrm{Q}}+(g-1)N_{\mathrm{Q}}=tN_0 + \chi
\end{equation}
where $t=ga$ and $\chi = (2g-ga-1)N_{\mathrm{Q}}$.
For cell $i+1$ it reads:
\begin{equation}
    N_{i+1} = tN_i +\chi =t^2N_{i-1} + t\chi+ \chi = t^3N_{i-2} + t^2\chi+ t\chi+\chi=...=t^{i+1}N_0 + \chi\sum_{k=0}^{i}t^k.
\end{equation}
We can thus evaluate the noise exiting the last cell $n_{\mathrm{c}}$ of the amplifier, relating the output noise $N_{n_{\mathrm{c}}} $ to the input noise $N_0$ 
\begin{equation}
    N_{n_{\mathrm{c}}} = TN_0 + \chi\frac{T-1}{t-1}
\end{equation}
where $T=\prod_{i=1}^{n_{\mathrm{c}}}t =t^{n_{\mathrm{c}}}$ is the total transmission.
For $T\gg1$ we can define the quantum efficiency $\eta$ as \cite{macklin2015}
\begin{equation}
\eta = \frac{T}{N_{n_{\mathrm{c}}}}=\frac{a g-1}{g-1}
\end{equation}
The JTWPA contains  $n_c = 2037$ cells~\cite{macklin2015} and at the sweet spot we observe a JTWPA gain $G=\num{120.3(1)}$.
\newpage

\section{JTWPA reflection and transmission}\label{supp:ReflectiveJTWPA}
\begin{figure}[h!]
\includegraphics{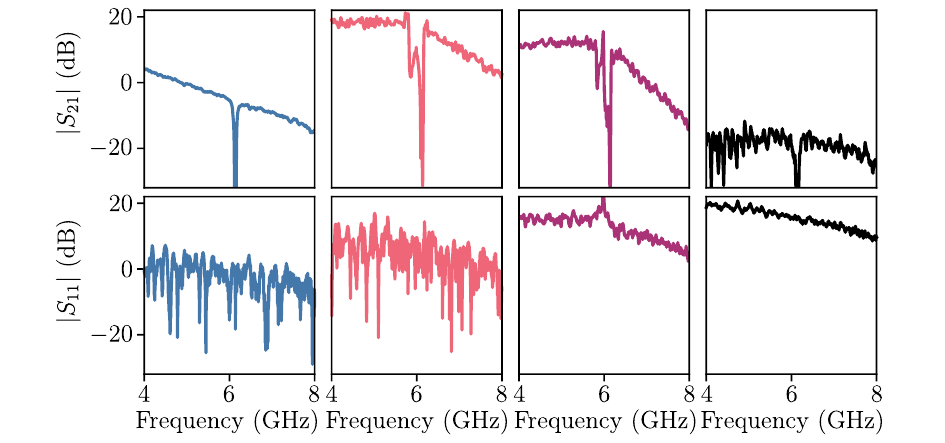}
\caption{JTWPA D1 transmission ($S_{21})$ and reflection ($S_{11}$) signals upon varying the pump power. Blue curve corresponds to the situation where no pump tone is applied, red purple and black correspond to a room temperature pump tone of $\qty{10}{\dBm}$, $\qty{12}{\dBm}$ and $\qty{18}{\dBm}$ respectively. 
}
\label{fig.twpa_reflection}

\end{figure}

We now show the JTWPA D1 (see main text) transmission and reflection upon varying the JTWPA pump power. These measurements are carried out as follows. We apply a pump tone at $\omega_{\mathrm{d}}/(2\pi) = \qty{5.985}{\giga\hertz}$ with a fixed power from nothing (blue curves in Fig.~\ref{fig.twpa_reflection}) to $\qty{18}{dBm}$ (black curves in Fig.~\ref{fig.twpa_reflection}). To measure the transmission and reflection profile we use a four port Virtual Network Analyzer (VNA) . We excite the JTWPA through the pump line (see Fig.~\ref{fig.fridge_setup}) and we then measure the two signal $S_{21}$ and $S_{11}$ resulting from passing through the JTWPA or being reflected.

We observe that just above the optimal point of operation (red curves in Fig.~\ref{fig.twpa_reflection}) the transmitted signal starts to degrade while the reflected signal increases (purple curves in Fig.~\ref{fig.twpa_reflection}). By going further in this non-linear regime (black curves in Fig.~\ref{fig.twpa_reflection}), we suppress transmission through the JTWPA.
\newpage

\section{BLAST optimization}\label{supp:BLAST}
\begin{figure}[h]
    \centering
    \includegraphics{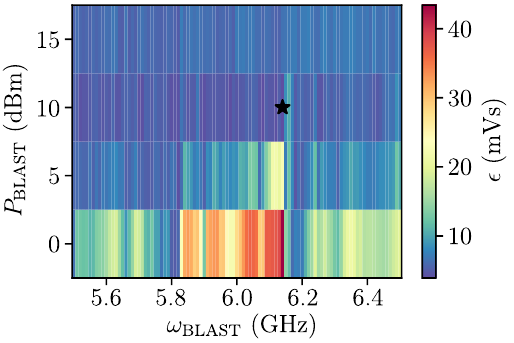}
    \caption{Integrated signal difference $\epsilon$ with and without BLAST tone as function of BLAST angular frequency and power. Black star corresponds to the optimal point of operation}
    \label{fig:blast}
\end{figure}

We show the optimization of the power and frequency of the BLAST tone. We have performed such experiment on JTWPA D1. We first measure a reference signal $I_{\mathrm{ref}}$ made of a flat-top pulse of length $\theta = \qty{0.15}{\micro\second}$ and rising time of $\qty{20}{\nano\second}$ with power of $\qty{-125}{\dBm}$ at the JTWPA input. To measure $I_{\mathrm{ref}}$, we operate the JTWPA in its sweet spot with a pump tone of angular frequency $\omega_{\mathrm{p}}/(2\pi) = \qty{5.985}{\giga\hertz}$ and power $\qty{-82}{\dBm}$ at the JTWPA input. We then perform the sequence shown in Fig.\ref{fig.4}a.  The three flat-top pulses have same pulse length $\theta=\qty{0.15}{\micro\second}$ and are delayed by $\tau=\qty{0.5}{\micro\second}$. The first and second pulse enter the JTWPA with power $\qty{-83}{\dBm},\qty{-78}{\dBm}$ respectively. In doing so, the third pulse, which is identical to $I_{\mathrm{ref}}$, is applied at the instant where we expect the dielectric echo to appear. The goal is to find a set of BLAST parameters that allows us to detect the small third pulse. We sweep the BLAST frequency and power to minimize the difference between the reference signal within the BLAST sequence $I_{\mathrm{ref}}^{\mathrm{BLAST}}$ and $I_{\mathrm{ref}}$. More precisely, we minimize the quantity 
\begin{equation}
\epsilon = \int_{t_0} ^{t_1}\left(I_{\mathrm{ref}}^{\mathrm{BLAST}}(t) - I_{\mathrm{ref}}(t)\right)dt
\end{equation}
In Fig.\ref{fig:blast} we remark that $\epsilon$ has a weak dependence on frequency and on $P_{\mathrm{BLAST}}$ power once it is strong enough to remove the dielectric echo. We find the BLAST optimal frequency to be $\omega_{\mathrm{BLAST}}/(2\pi)=\qty{6.14}{\GHz}$. The optimal power at the output of the local oscillator is $\qty{10}{\dBm}$ which corresponds to $\qty{-55}{\dBm}$ at the JTWPA input.

\section{Echo removal by interferometric cancellation}\label{supp:interferometry}
\begin{figure}
    \centering
    \includegraphics{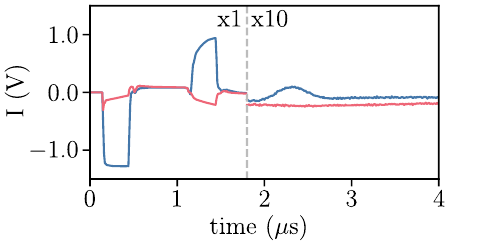}
    \caption{Detected I quadrature of a two pulse sequence as function of time. The two pulses have length $\theta = \qty{0.3}{\micro\second}$ and are separated by $\qty{1.0}{\micro\second}$. Lines correspond to detected signal when no compensation is applied (blue) and when it is applied (red).  }
    \label{fig:compensation_PC}
\end{figure}

The results in Fig.\ref{fig.4}(b) can be compared to a complementary strategy to remove JTWPA dielectric echoes, namely performing interferometric cancellation of the high power tones before the JWTPA input. We use a similar setup to the one presented in Fig.~\ref{fig.fridge_setup}. We apply concurrently a two pulse sequence made of rectangular pulses on the DUT (probe pulses) and on the JTWPA pump line (cancellation pulses). By tuning the amplitude, phase and delay of the cancellation pulses we can indeed have destructive interference between the two, preventing thus the excitation and consequent emission of an echo from the JTWPA dielectric defects, as shown in Fig.~\ref{fig:compensation_PC}
\end{document}